\documentstyle[prl,aps,twocolumn,epsfig]{revtex}

\begin{document}

\draft

\twocolumn[\hsize\textwidth\columnwidth\hsize\csname@twocolumnfalse\endcsname

\title{Ferromagnetic instabilities in atomically-thin lithium and sodium wires}
\author{A. Bergara$^{1,\ast}$, J. B. Neaton$^2$, and N. W. Ashcroft$^1$}
\address{
$^1$Laboratory of Atomic and Solid State Physics and the Cornell
Center for\\
Materials Research, Cornell University, Ithaca, NY 14853-2501, USA\\
$^2$ Department of Physics and Astronomy, Rutgers University, Piscataway, NJ 08855-0849, USA}
\date{December 4, 2001}
\maketitle


\begin{abstract}
Using density functional theory the ground state structural,
electronic, and magnetic properties of monatomic lithium
and sodium chains with low average density are investigated. A metallic, 
zigzag ground state structure is predicted but, most interestingly, 
stable equilibria for chains under tension
are predicted to be {\it ferromagnetic}, which can be traced to
exchange effects arising from occupation of the second subband as a 
function of the interatomic distance.
\end{abstract}



]


Substantial variations in structural, electronic, and magnetic
properties of solids have been predicted to occur with decreasing dimensionality,
and with the advent of new techniques in atomic manipulation and
nanofabrication\cite{eigler,uchida,salling}, it is now possible to realize quasi-one
dimensional systems experimentally. Using photoemission spectroscopy,
both Luttinger-liquidlike properties \cite{segovia}
and Peierls instabilities \cite{yeom} have
already been observed in linear monatomic chains formed on insulating substrates.
Despite confirmation of such unusual structural and electronic
properties, the existence of magnetic order in quasi-one-dimensional systems 
remains somewhat controversial\cite{thomas,malatesta,gold}.
Generalizing an earlier prediction of Paul\cite{paul}, the Lieb-Mattis
theorem\cite{lieb} formally prohibits ferromagnetic ground states
for strictly one dimensional systems in the absence of spin- or momentum-dependent forces.
Although strictly valid for single-band systems, this theorem is no longer applicable when two or more
transverse subbands are occupied, and in fact we report here 
the results of first-principles calculations
indicating a {\it ferromagnetic} ground state for lithium chains, as well as
stable magnetic equilibria for quasi-one dimensional sodium chains under tension.
The unexpected development of magnetic order in these classic nearly-free electron
systems  can be traced to exchange effects arising from 
the changing occupation of a second subband as a function of the interatomic distance.
Accordingly, at high \cite{na,bna} and now at low densities, lithium and sodium depart radically from the
simple-metallic behavior that typifies their normal density electronic structure.

To examine lithium and sodium chains over a wide range of interatomic distances,
we use a plane wave implementation of density functional theory \cite{hk} within the
local spin-density approximation\cite{ks}, the Vienna {\it ab initio
Simulation Package} (VASP) \cite{kresse}
Our Li and Na ultrasoft pseudopotentials\cite{vanderbilt} treat only the single outer $s$ electron
as valence; they also include non-linear core corrections\cite{froyen},
and both reproduce previously calculated structural and electronic properties of bulk
Li and Na. We use a supercell containing four atoms aligned along only one direction (defining the 
chain) and include 20 \AA\ of vacuum in the orthogonal directions to minimize interactions
between neighboring chains. For several different values of the chain-length per atom, $x$, we
optimize both the interatomic distance, $d$, and the bond-angle, $\alpha$; the resulting binding
energies appear in Fig. 1. As can be seen, the energy has a pronounced minimum in both cases
(around $x=1.7$ \AA\ for sodium and $x=1.55$ \AA\ for lithium),
at which point $d=3.47$ \AA \ and $\alpha=61^{\rm o}$ 
for sodium and $d=2.85$ \AA \ and $\alpha=63^{\rm o}$ for lithium.
Therefore we conclude that linear light-alkali chains are unstable to a 
lateral buckling into a planar zigzag or polymer-like configuration.
Out-of-plane deformations do not result in lower energies.
Previous experimental work by Whitman {\it et al.}\cite{whitman}
observed that monatomic cesium chains on GaAs and InAs substrates form a zig-zag structure,
which then remained stable under tension. Recent first-principles calculations
have also suggested a similar zigzag structure in monatomic gold chains \cite{sanchez};
a similar zigzag geometry was proposed\cite{march} to explain expanded
liquid Rb and Cs.

Although both alkali chains satisfy the necessary condition for insulating behavior with
two monovalent atoms per unit cell, band overlap nevertheless prevents the systems from being
insulating. When perfectly linear, the electronic structure of the each chain consists
of a single half-filled $s$-like band, with two degenerate, unoccupied $p$-like transverse subbands lying above.  
The electronic structure of the {\it zigzag} chain, however, 
exhibits {\it two} partially occupied subbands, one predominately $s$-like and one $p$-like.
Formation of the zigzag structure lifts the two-fold degeneracy of the second subband 
(corresponding to the two transverse $p$-like bands of the linear chain), 
one of which then drops below the Fermi energy, stabilizing the zigzag structure over the linear chain. 
This symmetry-breaking also increases the coordination of each atom from two
to four nearest-neighbors, lowering Madelung-type terms and further stabilizing
the zigzag arrangement. Additionally, we note that distances between neighbors in both chains are 
longer than the bond length of the molecule (2.99 \AA \ for sodium and 2.71 \AA \ for lithium) but smaller than 
the shortest interatomic distance in their equilibrium BCC solid phases (3.66 \AA \ and 2.92 \AA \, respectively), 
reflecting an expected increase in bond length with coordination number.

\begin{figure}[t!]
\epsfig{figure=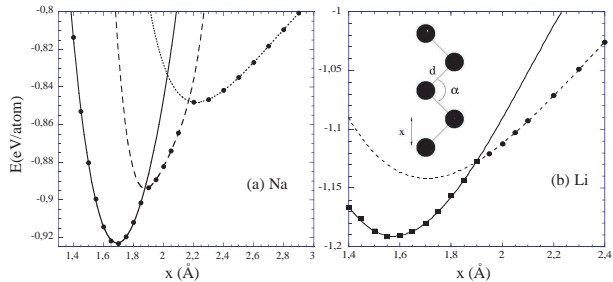,width=3.75cm,angle=90}
\vskip 2mm
\caption{Energy per atom (solid symbols) plotted as function of the chain-length per atom, $x$,
for (a) sodium and (b) lithium chains.
The energy in each regime for both chains is well
fit to $E(x)=a_0+a_1x+a_2x^2+a_3x^3+a_4x^4$ (solid, dashed and dotted
lines). For each value of $x$, the ions are relaxed into their equilibrium
positions until forces calculated from
the Hellman-Feynman theorem are less than 1 meV/\AA. In all calculations
we use a $30\times 3 \times 3$ {\bf k}-mesh and plane wave cut-off of 20
Ry, together resulting in a total energies
converged to 1 meV/atom. When converging to the ground state, 
fractional occupancies are treated using a Gaussian smearing
of $\sigma=0.1$ eV. The inset shows the three parameters
characterizing the geometry of the chains: $x$ (the chain-length per atom), $d$ (the bond-length), and $\alpha$ (the
bond-angle).}
\label{1}
\end{figure}

Along with the energy minimum at a length of $x=1.7$, the energy of the sodium chain has two
noticeable kinks (at $x=1.9$ and 2.2 \AA ). Similar features are found in the lithium chain, 
though only one kink is observed (at $x=1.95$ {\AA}) upon stretching (Fig. 1(b)).
In both chains,  we record complementary
discontinuous changes in both $d$ and $\alpha$, each associated with major atomic rearrangements 
(as will be illustrated in Fig. 4 below).
The additional metastable equilibria observed in sodium exhibit surprising
electronic and magnetic properties.
As $x$ increases, the second subband becomes progressively less occupied, and this 
has significant consequences for the emerging magnetic properties, as we now show.
(Although in the following our discussion will focus primarily on the 
sodium chains, our calculations indicate that equivalent conclusions are obtained for lithium.)
The band structure corresponding to sodium chains for $x=2$ \AA \ is shown in Fig. 2 and, remarkably,
the bands for electrons with opposite spins are shifted with respect to one
another, indicating that the sodium chains have {\it a non-zero net
spin polarization}. 

Interestingly the lithium chain already has a net spin polarization at its
equilibrium stabilizing length, thereby accounting for the absence of the first kink in Fig. 1(b). 
The spin-split $p$-like band,
corresponding to antisymmetric orbitals oriented transverse to the
direction of the chain, allows for electrons of a single spin
character only and is thus solely responsible for the magnetic polarization of the
chain. The occupation number corresponding to this subband is so low ($\sim
0.15$\ $e$/atom) that the exchange energy actually overcomes the excess of kinetic and correlation energies, 
resulting in the ferromagnetic state. We further note that the 
density of states (DOS) is observed to vary roughly as  $1/\sqrt{E-\epsilon^\sigma_m}$
(where $\epsilon^\sigma_m$ is the energy of subband $m$ with spin $\sigma$ at the $\Gamma$ point), 
as is expected
for free electrons in one dimension. This behavior arises from van Hove
singularities at the bottom of each subband, which are especially prominent in one
dimension. We note in passing that although our calculations carried out within the generalized
gradient approximation (GGA) \cite{perdew} result in a slightly larger
interatomic distance $d$, they do not change the overall notably
``polymeric'' structure of the chains. The closing of the
second subband and corresponding transition to the nonmagnetic state
is found to occur at the same chain length, but the
magnetic instability remains for a larger range of chain lengths within the GGA.

\begin{figure}[t!]
\center
\epsfig{figure=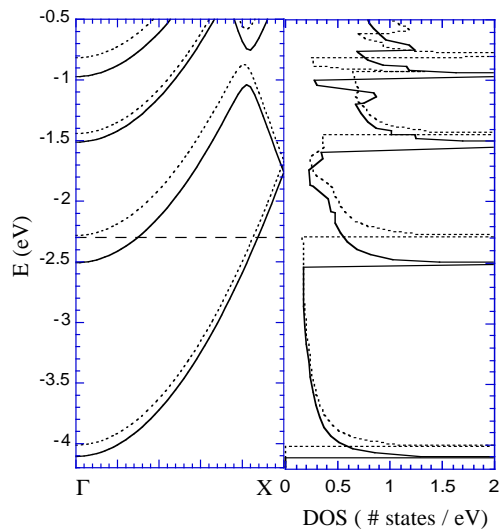,width=7cm,angle=0}
\vskip 2mm
\caption{(a) Band structure of the optimized sodium chain at $x=2$ \AA. 
Bands of opposite spin are denoted by the dotted lines.
The Fermi energy (dashed horizontal line) is located at the bottom
of the second spin-down subband. (b) Spin-resolved
density of states (DOS). As expected for free electrons in one dimension, the
contribution to the DOS of each subband follows $1/\sqrt{E-\epsilon^\sigma_m}$,
$\epsilon^\sigma_m$ being the energy
of the bottom of subband $m$ with spin $\sigma$.}
\label{2}
\end{figure}

The valence charge and spin polarization for sodium chain with $x=2.0$
\AA\ appear together in Fig. 3. Notice that the charge is mainly confined to the region between the
ions, and is almost uniform along the longitudinal direction of the chain. There does not
appear to be any directional bonding between the ions: covalent bonding will therefore
not account for the stability of the zigzag structure. The minima on either side of
the charge density maxima correspond to the location of the ions, where the valence charge density
is small because of the pseudopotential approximation.  Because
of the $p$-symmetry of the second subband, the spin density also exhibits interesting
character: in Fig 3(b) {\it two separate maxima} can be seen on both sides with a minimum in the center.
The spin polarization in this valley has opposite sign as compared to the net magnetic moment. The
magnetic polarization of the chain is therefore highly anisotropic 
and, as previously indicated, the chain is
ferromagnetic along $x$ but has an {\it antiferromagnetic} character 
along the {\it transverse} direction of the chain.
Similar charge and spin densities can be observed for lithium zigzag chains.

\begin{figure}[t!]
\center
\epsfig{figure=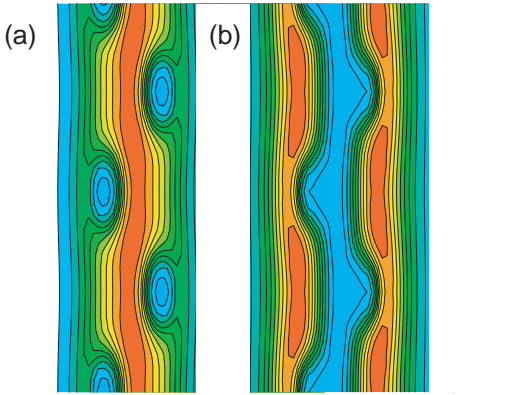,width=7cm,angle=0}
\vskip 2mm
\caption{Valence charge density $\rho_{\rm up} + \rho_{\rm down}$ (a)
and spin density $\rho_{\rm up} - \rho_{\rm down}$ (b) of
the sodium zigzag chain at $x= 2 $ \AA \ as determined within the LSDA.
The valence charge density maxima (red areas) extend along the chain between the ions, 
 where the charge density is minimal (concentrated blue areas) owing to the pseudpotential
approximation. In (b), oblate red regions of large spin density are 
located on both sides of the chain, away from the central charge density minima (in blue-green).
}
\label{3}
\end{figure}

When the chain length is extended to $x=2.2$ \AA \,, the second subband 
rises above the Fermi energy and the gap between spin-split subbands
closes; the ferromagnetic order vanishes,
resulting in the final kink observed in Fig. 1(a) for the sodium chain.  
We may also note that for the paramagnetic state (having only one occupied subband) 
the energy is well fit to a polynomial function of $x$ (as in Fig. 1), and the minimum energy
obtained from this fit, corresponding to the cohesive energy of the
chain with only the first subband occupied, is higher than the value
calculated for the ground state of the chain.
Thus filling the second band decreases the total energy and
stabilizes the zigzag chain. The net magnetic moment per atom and the
optimal geometry of the zigzag chains as a function of $x$ for both
alkali chains is shown in Fig. 4.  The maximum value of
the net magnetic moment of the chain (per conduction
electron) is $\sim 0.18 \mu_B$ ($\sim 0.16 \mu_B$ for lithium),
and it decreases with increasing $x$; and for $x>2.2$ \AA, only
one subband is occupied and the chain is once again paramagnetic.
The discontinuities in $d$, corresponding to the kinks observed for the
energy and major atomic rearrangements, can be seen to
reflect the onset of magnetic transitions.

According to our results, the spontaneous magnetization in
these zigzag chains originates from the exchange between electronic states
in the second subband; these results are therefore not in violation of
the Lieb-Mattis theorem, which holds strictly 
for systems in which only a single subband is occupied.
Recent calculations\cite{nerea} have shown that sodium quantum wires,
described within the jellium model, undergo spontaneous magnetization
for certain radii.  In particular, they have predicted that the ground state
is fully spin polarized when only the lowest subband is occupied. The
effects of band structure and associated orbital hybridization, absent from the jellium
model, clearly have an important influence on
the stability of the chain. When the electron-ion interaction and
atomic structure are explicitly considered, we have found that the
ground state of the chain is paramagnetic when only the second subband is occupied. 
At larger chain lengths (i.e., $x>2.2$ \AA\ ) the first
subband is completely filled and although the Fermi wave vector makes
contact with the edge of the Brillouin zone, we do not observe a
Peierls transition. Instead, we find an intermediate state where the
chain remains metallic and paramagnetic until  $x=3.5$ \AA\,, above which
dimerization drives the metal-insulator transition.

\begin{figure}[t!]
\center
\epsfig{figure=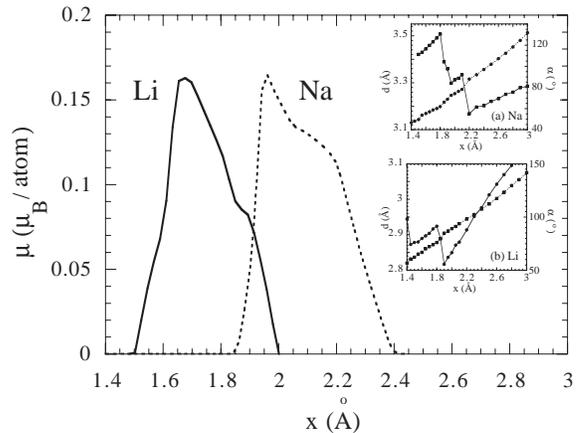,width=9cm}
\vskip 2mm
\caption{The magnetic moment per atom
as a function of the length per atom, $x$, for both lithium (solid)
and sodium (dashed) chains. Lithium chains exhibit a stable ferromagnetic
{\it ground state}; ferromagnetism develops in sodium chains only under tension.
The inset shows the optimized bond-angle, $\alpha$ (solid line), 
and bond-length, $d$ (dashed line), plotted as
a function of the chain-length per atom, $x$. }
\label{4}
\end{figure}

In summary, we observe that one-dimensional lithium and sodium chains are unstable
to zigzag polymer-like structures. The transition to the zigzag structures increases the coordination number 
and lowers the energy of a minimally-occupied $p$-like subband (possessing a large van Hove peak). 
The spin-splitting of a $p$-like subband moves the van Hove peak to lower energies in lithium,
and a ferromagnetic ground state is predicted;
for sodium, this ferromagnetic ordering is stable only when the chains are under tension.
Our calculations clearly establish an intimate connection between 
the existance of magnetic order and zigzag arrangements. But the 
fact that magnetic order is found at all in these structures raises the
important possibility of the occurance of more general ground states,
such as, for example, non-collinear magnetic order, or even static spin waves 
as originally proposed for the alkalis by Overhauser \cite{overhauser}.
Accordingly, future extension of this work along these lines would be of considerable interest.
The results presented here for lithium and sodium
may not necessarily be extended to chains composed of heavier elements. 
For the sake of comparison, monoatomic gold chains also show
the same zigzag structures\cite{sanchez} and the energy shows two
minima with the intermediate maximum associated to the closing of a conduction band.
However, we have found a very small magnetic instability in gold chains,
$\mu_{\rm max}\sim 0.02 \mu_B$, and we believe that is related
to the presence of $d$ bands near the Fermi energy. 
The possibility of
magnetic chain- or stripe-like structure is also predicted to arise 
in an important system which might be seen to underlie the 
alkali chains considered here, namely the 
two-dimensional interacting electron gas at low densities \cite{bonev}.
Future experiments on inert substrates (e.g., cleaved rare-gas crystals), 
chosen so that they not interfere with
the $p$-like subbands, would clearly be of very significant interest.

This work was supported by the National Science Foundation (DMR-9988576).
This work made use of the
Cornell Center for Materials Research Shared Experimental Facilities,
supported through the
National Science Foundation Materials Research Science and Engeneering
Centers program (DMR-0079992).
A.B. would like to acknowledge financial support from the Spanish
Ministerio de Educaci\'on y
Cultura under Fulbright Grant No. FU-99-30656084.

\end{document}